\documentclass[a4paper,11pt]{article}
\usepackage[toc,page]{appendix}
\usepackage{graphicx}
\usepackage{colortbl}
\usepackage{amsmath}
\usepackage{subfigure}
\usepackage{mathtools}

\DeclarePairedDelimiterX\braket[2]{\langle}{\rangle}{#1 \delimsize\vert #2}
\usepackage[utf8]{inputenc}
\usepackage{float}
\usepackage[normalem]{ulem}
\usepackage{epstopdf}
\usepackage{amsfonts,amsmath,amssymb}
\usepackage{hyperref}

\usepackage{epsfig,multicol,bbm}
\usepackage{color}
\usepackage[dvipsnames]{xcolor}

\definecolor{darkblue}{rgb}{0.0, 0.0, 0.55}
\definecolor{grey}{rgb}{0.57, 0.64, 0.69}
\definecolor{lightbrown}{rgb}{0.71, 0.4, 0.11}

\newcommand{\be}{\begin{equation}}
\newcommand{\ee}{\end{equation}}

\newcommand\fverb{\setbox\pippobox=\hbox\bgroup\verb}
\newcommand\fverbit{\egroup\item[\fbox{\unhbox\pippobox}]}

\newbox\pippobox
\begin{document}
\title{\bf Holographic Entanglement Entropy in $T\bar{T}$-deformed CFTs}
\author{ M. R. Setare\thanks{Electronic address: rezakord@ipm.ir}\,,\,S. N. Sajadi\thanks{Electronic address: naseh.sajadi@gmail.com}
\\
\small Department of Science, Campus of Bijar, University of Kurdistan, Bijar, Iran\\
%\small Physics Department and Biruni Observatory, College of Sciences, Shiraz University,
%Shiraz 71454, Iran
}
\maketitle
\begin{abstract}
In this work, we study the holographic entanglement entropy of two dimensional $T\bar{T}$-deformed conformal field theory. We compute the correction due to the deformation up to
the leading order of the deformation parameter in the framework of New massive gravity, General minimal massive gravity, and Exotic general massive gravity. We also apply Song's prescription to obtain the entanglement entropy for deformed theories. In each cases, we find agreement between the results.
\end{abstract}

\maketitle
\section{Introduction}

Holographic conjecture is one of the powerful tools to study quantum gravity, in which the
quantum gravity in $d$ spacetime dimension is equivalent to a quantum field theory on $d-1$
dimensions boundary. An important example is a holographic duality between conformal
field theory in $d$ dimension and $(d+1)$-dimensional AdS gravity \cite{hol01}-\cite{hol03}.
Conformal field theory is by definition a UV complete framework, in which the rules of
local quantum field theory apply at all energy scales.
CFTs, are critical points of RG flows. It is then natural to ask: can holography be extended to effective QFTs for which the UV behavior is not described by a CFTs? In the context of AdS$_3$/CFT$_2$, this the question has been answered by Zamolodchikov \cite{Smirnov:2016lqw} by considering a general class of exactly solvable irrelevant deformations of 2D CFT. In \cite{McGough:2016lol} proposed that in the holographic dual, this deformation represents a geometric cut-off on a wall at finite radial distance $r = r_c$ in the bulk that removes the asymptotic region of AdS and places the QFT on it. More precisely if a CFT has a gravity dual, then the deformed theory is dual to the original gravitational theory with the new boundary at $r = r_c$. In this paper following the papers \cite{Chen:2018eqk} and \cite{Allameh:2021moy} we would like to understand the effect of this deformation on entanglement entropy using holography in the framework of higher derivative massive gravity in $2+1$ dimensions.\\ Because of the absence of local degrees of freedom, General Relativity (GR) in three dimensions is an easier theory for studying the different aspects of gravity. New Massive Gravity(NMG) is a three-dimensional theory of gravity with parity-even, higher derivative action which at the linearized level reduces to massive spin-two Fierz-Pauli theory \cite{Bergshoeff:2009hq}, \cite{Bergshoeff:2009aq}.\\ {General Minimal Massive Gravity (GMMG) which was introduced in \cite{Setare:2014zea}, is an example of the 3D theory of gravity with actions that makes use of two auxiliary one-forms, $h$ and $f$, which at the level of the field equations can be integrated out, leading to the New Massive Gravity field equations supplemented by the Cotton tensor and by a parity even tensors, $J_{a b}$. This term with respect to the curvature is quadratic, and therefore the field equations for the metric remain of fourth-order. These effective Einstein equations cannot be obtained only from a variational principle of the metric as a dynamical field, nevertheless, they are on-shell consistent as it is the case in the theories introduced in \cite{Bergshoeff:2014pca}, \cite{Tekin:2015rha}, \cite{Ozkan:2018cxj} and \cite{Alkac:2018eck}. GMMG avoids the bulk-boundary clash and therefore possesses positive energy excitations around the maximally AdS$_3$ vacuum as well as a positive central charge in the dual CFT. Such problem in the previously constructed gravity theories with local degrees of freedom in 2+1-dimensions namely Topologically Massive Gravity and the cosmological extension of New Massive Gravity is present \cite{Bergshoeff:2009hq}, \cite{Deser:1982vy}, \cite{Deser:1981wh}. Exotic general massive gravity is another 3D theory of gravity with parity$-$odd action which describes a propagating massive spin$-$2 fields. The field equations of this theory supplement the Einstein equations with a term that contains up to 3rd of the metric and is built with combinations and derivatives of the Cotton tensor \cite{Ozkan:2018cxj}. The different aspects of this model have been studied in \cite{Mann:2018vum}, \cite{Bergshoeff:2019rdb}, \cite{Giribet:2019vbj}, \cite{Setare:2021gll}, \cite{Setare:2021ugr}, \cite{Setare:2021ref}.}

The paper is organized as follows: In section \ref{sec2}, we obtained the entanglement entropy for NMG with parity even action, directly by using on-shell action and using the RT-method. In section \ref{sec3}, for the other Chern-Simons like theories of gravity GMMG and EGMG we obtain the entanglement entropy and repeat the procedure of previous section for them. We provide some conclusions in Section \ref{sec4}.

\section{Entanglement Entropy for NMG}\label{sec2}
Among massive gravity models in three dimensions, a well known one is that of the new massive gravity. This model is equivalent to the three-dimensional Fierz-Pauli action for a massive spin$-2$ field at the linearized level. In addition, NMG preserves parity symmetry which is not the case for the topological massive gravity (TMG).
The action of NMG is presented as follows
\begin{equation}\label{action}
S_{NMG}=\dfrac{1}{8\pi G}\int d^{3}x\sqrt{-g}\left[R-2\lambda -\dfrac{1}{m^2}(R^{\mu \nu}R_{\mu \nu}-\dfrac{3}{8}R^{2})\right]
\end{equation}
where $\lambda$ and $M$ are the cosmological constant and the parameter of NMG, respectively.
By a variation of the Lagrangian we obtain
\begin{equation}\label{eq1}
E_{\mu \nu}=G_{\mu \nu}+\lambda g_{\mu \nu}-\dfrac{1}{2m^{2}}N_{\mu \nu},
\end{equation}
with
\begin{align}\label{eq2}
N_{\mu \nu}=-\dfrac{1}{2}\nabla^{2}R g_{\mu \nu}-\dfrac{1}{2}\nabla_{\mu}\nabla_{\nu} R+2\nabla^{2}R_{\mu \nu}+4 R_{\mu a \nu b}R^{a b}-
\dfrac{3}{2}R R_{\mu \nu}-R_{\alpha \beta}R^{\alpha \beta}g_{\mu \nu}+\dfrac{3}{8}R^{2}g_{\mu \nu},
\end{align}
and $G_{\mu \nu}$ is the Einstein tensor.
To obtain the renormalized action we should add the generalized Gibbons-Hawking boundary term for NMG and the counter-term to the action (\ref{action}) as follows
\begin{equation}\label{GHYbound}
S_{GH}=\int d^{2}x\sqrt{-\gamma}\left(2 K+\hat{f}^{a b}K_{a b}-\hat{f}K\right),\;S_{ct}=\left(1+\dfrac{1}{2m^2 l^2}\right)\int d^{2}x \sqrt{-\gamma}(1+\delta^2 \kappa(z,\bar{z})).
\end{equation}
where
\begin{align}
K_{\mu \nu}&=-\dfrac{1}{2}(n_{\mu ;\nu}+n_{\nu ;\mu}),\;\gamma^{\mu \nu}=g^{\mu \nu}-n^{\mu}n^{\nu},\;\hat{f}^{a b}=f^{\mu \nu}\gamma_{\mu}{}^{a}\gamma_{\nu}{}^{b}\;
f_{\mu \nu}=\dfrac{2}{m^2}\left(R_{\mu \nu}-\dfrac{1}{4}R g_{\mu \nu}\right).
\end{align}

Now, let us consider a deformed CFT on manifold $\mathcal{M}$. The entanglement entropy is given by
\begin{equation}
S_{EE}=\lim_{n\to 1}\dfrac{1}{1-n}\log\frac{Z_{n}}{Z^{n}}
\end{equation}
where $Z_{n}$ is the partition function on $\mathcal{M}^{n}$ which is obtained by using the replica method in which one should
prepare $n$ copies of the manifold on which the original field theory lives and gluing them together.

We start with the deformed CFT defined on the boundary metric in two dimensions with complex coordinates ($x,\bar{x}$) to cover this surface as
\begin{equation}
ds^2=dx d\bar{x}.
\end{equation}
By using the following transformation of the coordinates one can convert the metric to a conformal form as
\begin{equation}
y=\left(\dfrac{x-a}{x-b}\right)^{\frac{1}{n}}
\end{equation}
then
\begin{equation}\label{eqbound}
ds^2=e^{\phi(y,\bar{y})}dy d\bar{y},\;\;\;\;\;\; e^{\frac{\phi(y,\bar{y})}{2}}=nl \dfrac{\vert y\vert^{n-1}}{\vert y^{n}-1\vert^2}
\end{equation}
where $\phi$ is the Liouville field. One can construct the bulk spacetime
associated to the boundary metric (\ref{eqbound}) by using the Fefferman-Graham metric \cite{f1}
\begin{equation}
ds^2=\dfrac{d\rho^2}{4\rho^2}+\dfrac{1}{\rho}g_{ij}(\rho,X)dX^{i}dX^{j},\;\;\;\; g_{ij}(\rho, X)=g_{ij}^{(0)}(X)+\rho g_{ij}^{(1)}(X)+...
\end{equation}
here $X^{i} = (y,\bar{y})$ and $g^{(0)}_{ij}=e^{\phi}dyd\bar{y}$. So the bulk metric can be written as follows \cite{f2}
\begin{equation}
ds^2=\dfrac{d\rho^2}{4\rho^2}+\dfrac{1}{\rho}e^{\phi}dyd\bar{y}+\dfrac{1}{2}T_{\phi}dy^2+\dfrac{1}{2}\bar{T}_{\phi}d\bar{y}^2+\dfrac{1}{4}R_{\phi}dyd\bar{y}+\dfrac{1}{4}\rho e^{-\phi}(T_{\phi}dy+\dfrac{1}{4}R_{\phi}d\bar{y})(\bar{T}_{\phi}d\bar{y}+\dfrac{1}{4}R_{\phi}dy)
\end{equation}
where
\begin{equation}
R_{\phi}=4\partial_{y}\bar{\partial}_{y}\phi,\;\;\;\; T_{\phi}=\partial^{2}_{y}\phi -\dfrac{1}{2}(\partial_{y}\phi)^2,\;\;\;\;\; \bar{T}_{\phi}=\bar{\partial}^{2}_{y}\phi -\dfrac{1}{2}(\bar{\partial}_{y}\phi)^2.
\end{equation}
The following coordinate transformations \cite{f3}
\begin{equation}
\xi=\sqrt{\dfrac{e^{\phi}}{\rho}}+\dfrac{1}{4}\sqrt{\dfrac{\rho}{e^{\phi}} }\vert \partial_{y}\phi\vert^2 ,\;\;\;\; z=y+\dfrac{1}{2}\dfrac{\rho e^{\phi}\bar{\partial}_{y}\phi}{1+\dfrac{1}{4}\rho e^{-\phi}\vert \partial_{y}\phi \vert^{2}}
\end{equation}
convert the FG coordinate to the Poincare coordinate, which brings us to the following metric
\begin{equation}\label{eqmetric}
ds^2=\dfrac{d\xi^2}{\xi^2}+\xi^2 dz d\bar{z}.
\end{equation}
This metric is a solution for NMG if
\begin{equation}
\lambda=-\left(1+\dfrac{1}{4m^2}\right).
\end{equation}
 The NMG on-shell action is given by
\begin{equation}
S_{NMG}=\left(1+\dfrac{1}{2m^2}\right)\left[-\dfrac{1}{\delta^2}\iint e^{\phi}dzd\bar{z}-\dfrac{1}{2}\iint\psi dzd\bar{z}-\dfrac{\delta^2}{16}\iint e^{-\phi}\psi^2 dzd\bar{z}\right]
\end{equation}
here we assumed $\rho=\delta^2$, then the regulator surface is
\begin{equation}
\xi_{f}=\dfrac{1}{\delta}e^{\frac{\phi}{2}}+\dfrac{\delta}{4}e^{-\frac{\phi}{2}}\psi.
\end{equation}
While on the boundary $n^{\mu}=\xi \delta^{\mu}_{\xi}, \gamma=-\xi^4/4, g=-\xi^2/4$ and $K=\gamma^{a b}K_{a b}=2$ one can get the Gibbons-Hawking term and the counter term as
\begin{equation}
S_{GH}=\left(1+\dfrac{1}{2m^2}\right)\left[\dfrac{2}{\delta^2}\iint e^{\phi}dzd\bar{z}+\iint \psi dzd\bar{z}+\dfrac{\delta^2}{8}\iint e^{-\phi}\psi^2 dzd\bar{z}\right],
\end{equation}
and
\begin{equation}
S_{ct}=-\left(1+\dfrac{1}{2m^2}\right)\left[\dfrac{1}{\delta^2}\iint e^{\phi}dzd\bar{z}+\iint \dfrac{1}{2}\psi +\kappa(z,\bar{z})e^{\phi(z,\bar{z})} dzd\bar{z}+\dfrac{\delta^2}{2}\iint \left(\dfrac{1}{8}e^{-\phi}\psi^2 +\kappa(z,\bar{z})\psi\right) dzd\bar{z}\right].
\end{equation}
Thus, by choosing $\kappa=\psi/8e^{\phi}$ one can get the renormalized on-shell action is given by
\begin{equation}
S_{r}=S_{NMG}+S_{GH}+S_{ct}=-\dfrac{1}{8}\left(1+\dfrac{1}{2m^2}\right)\iint\left[\psi+\dfrac{\delta^2}{2}\psi^2 e^{-\phi}\right]dzd\bar{z}.
\end{equation}
It is possible to rewrite the renormalized action in the following form
\begin{equation}
S_{r}=\dfrac{1}{64\pi G}\left(1+\dfrac{1}{2m^2}\right)\int dV\left(\partial_{i}\phi\partial^{i}\phi +\dfrac{\delta^2}{2}\left(\partial_{i}\partial^{i}e^{\frac{\phi}{2}}\right)^2\right),
\end{equation}
after integrating by part one can get
\begin{equation}
S_{r}=\dfrac{1}{64\pi G}\left(1+\dfrac{1}{2m^2}\right)\int dS_{n}\left[\phi \partial_{n}\phi +\dfrac{\delta^2}{2}\left(\partial_{n}e^{-\frac{\phi}{2}}\square e^{-\frac{\phi}{2}}-e^{-\frac{\phi}{2}}\partial_{n}\square e^{-\frac{\phi}{2}}\right)\right].
\end{equation}
To evaluate this integral we adopt
\begin{equation}\label{eqq23}
z=re^{i\theta},\;\;\;\;\;\;\; e^{-\frac{\phi}{2}}=\dfrac{1}{n\ell}(r^{n+1}+r^{-n+1}-2r\cos(n\theta)),\;\;\;\;\; e^{-\phi}\psi^2=(\partial \bar{\partial}e^{-\frac{\phi}{2}})^2.
\end{equation}
Then one can get
\begin{equation}
S_{EE}=\dfrac{1}{4G}\left(1+\dfrac{1}{2m^2}\right)\left(\dfrac{1-n^2}{n}\right)\left[\log\left(\dfrac{\ell}{\delta}\right)+\dfrac{\delta^2}{n\ell^2}\right],
\end{equation}
one finally arrives at
\begin{equation}\label{eqqentropy1}
S_{EE}=\dfrac{c}{3}\log\left(\sqrt{\dfrac{24\pi}{\varsigma c}}\ell\right)+\dfrac{\varsigma c^2}{72 \pi \ell^2},
\end{equation}
where
\begin{equation}
\varsigma=\dfrac{8Gm^2\delta^2}{1+2m^2}.
\end{equation}
In the limit $m\to \infty$, $\varsigma=4\pi G\delta^2$.
It should be notice that the central charge of  the deformed CFT is the same as the original CFT.\\ It is expected that the holographic entanglement entropy of deformed CFT is also given by the RT-formula.
By using $z=x+\tau$, $\xi=1/\eta$ and going to the polar coordinate in \eqref{eqmetric}, we have
\begin{equation}\label{eqqpon}
ds^2=\dfrac{1}{\eta^2}\left[d\eta^2+r^2d\tau^2+n^2dr^2\right]
\end{equation}
The entanglement entropy can be calculated using the presymplectic potential near the special curve by replacing $\delta g_{\mu \nu}=\partial_{n}g_{\mu \nu}$ as
\begin{equation}
S_{HEE}=\int_{\gamma_{A}\times S^{1}}\Theta\left(\phi_{i},\partial_{n}\phi_{i}\right)_{n\to 1, r\to 0},
\end{equation}
which is a surface integral on $\gamma_{A}$ after integrating out the $\tau$ along the $S^{1}$.
For NMG the presymplectic potential is given by
\begin{align}\label{pres}
\Theta^{\mu}_{NMG}&=\theta^{\mu}-\dfrac{1}{2}f \theta^{\mu}+f^{\rho \sigma}g^{\mu \nu}\bar{\nabla}_{\rho}(\delta g_{\sigma \nu})-\dfrac{1}{2}f^{\rho \sigma}\bar{\nabla}^{\mu}(\delta g_{\rho \sigma})-\dfrac{1}{2}f^{\mu \nu}g^{\rho \sigma}\bar{\nabla}_{\nu}(\delta g_{\rho \sigma})\nonumber\\
&+\dfrac{1}{2}[\bar{\nabla}^{\mu}f^{\nu \rho}-2\bar{\nabla}f^{\mu \rho}+g^{\mu \nu}\bar{\nabla}^{\rho}f+g^{\nu \rho}\bar{\nabla}_{\sigma}f^{\sigma \mu}-g^{\nu \rho}\bar{\nabla}^{\mu}f]\delta g_{\nu \rho},\;\;\;\; \nonumber\\
&\theta^{\mu}=g^{\mu \nu}\bar{\nabla}^{\rho}(\delta g_{\nu \rho})-g^{\rho \sigma}\bar{\nabla}^{\mu}(\delta g_{\rho \sigma}).
\end{align}
 So, the presymplectic structure for metric \eqref{eqqpon} is obtained as
 \begin{equation}\label{eqpres}
\Theta^{r}=\dfrac{1}{8\pi G}\left(1+\dfrac{1}{2m^2}\right)\dfrac{\eta^2}{r n^3}.
\end{equation}
 Then, one can achieve the entropy as
 \begin{equation}\label{shees}
 S_{HEE}=\int_{0}^{2\pi} d\tau \int \sqrt{-g}\Theta^{r}dr=\dfrac{1}{4G n^2}\left(1+\dfrac{1}{2M^2}\right)\int_{\eta_{f}}^{\ell}\dfrac{d\eta}{\eta}=\dfrac{1}{4G}\left(1+\dfrac{1}{2m^2}\right)\ln\left(\dfrac{\ell}{\eta_{f}}\right),
 \end{equation}
here we have used $\sqrt{- g}=\frac{n r}{\eta^3}$, $\eta_{f}=1/\xi_{f}$ and $\ell$ is the interval length of subsystem $\mathcal{A}$.
In the case of $\delta\ll 1$, one can obtain
\begin{equation}\label{eqqhee}
S_{HEE}= \lim \limits_{\substack{%
    n \to 1\\
    r \to 0}}\dfrac{1}{4G}\left(1+\dfrac{1}{2m^2}\right)\left[\ln\left(\dfrac{\ell e^{\frac{\phi}{2}}}{\delta}\right)+\dfrac{\psi}{4e^{\phi}}\delta^2\right].
\end{equation}
 By using \eqref{eqq23} one can arrive
\begin{equation}\label{eqlimit}
 \lim \limits_{\substack{%
    n \to 1\\
    r \to 0}} e^{-\frac{\phi}{2}}=\dfrac{1}{\ell},\;\;\;\;\;\; \lim \limits_{\substack{%
    n \to 1\\
    r \to 0}} e^{-\phi}\psi=\dfrac{2}{\ell}.
\end{equation}
Inserting \eqref{eqlimit} into the \eqref{eqqhee}, it arrives \eqref{eqqentropy1}.

\section{Entanglement Entropy for GMMG}\label{sec3}
The Lagrangian of GMMG is obtained by generalization the Lagrangian of generalized massive gravity.
The Lagrangian of GMMG model is \cite{Setare:2014zea}
\begin{small}
\begin{equation}\label{eqlag}
L_{GMMG}=-\sigma e.R+\dfrac{\Lambda_{0}}{6}e.e\times e+h.T(\omega)+\dfrac{1}{2\mu}\left(\omega.d\omega+\dfrac{1}{3}\omega.\omega\times \omega \right)-\dfrac{1}{m^2}\left(f.R+\dfrac{1}{2}e.f\times f\right)+\dfrac{\alpha}{2}e.h\times h \ .
\end{equation}
\end{small}
Here $m$ is the mass parameter of NMG term, $h$ and $f$ are auxiliary one-form fields,
${\Lambda}_{0}$ is a cosmological parameter with dimension of mass squared, $\sigma$ is a convenient sign, $\mu$ is a mass parameter of Lorentz Chern$-$Simons term, $\alpha$ is a dimensionless parameter, $e$ is a dreibein and $\omega$ is a dualized spin-connection.
Once the auxiliary one$-$form fields $f$ and $h$ are integrated out, the field equations can be written as
\begin{equation}\label{eqfield}
\bar{\sigma}G_{\mu\nu}+\bar{\Lambda}_0g_{\mu\nu}+\frac{1}{\mu}C_{\mu\nu}+\frac{\gamma}{\mu^2}J_{\mu\nu}+\frac{s}{2m^2}K_{\mu\nu}=0\ ,
\end{equation}
where $C_{\mu\nu}$ is the Cotton tensor, $K_{\mu\nu}$ is the Euler$-$Lagrange derivative of the quadratic part of the NMG Lagrangian with respect to the metric, and $J_{\mu\nu}$ is the quadratic in the curvature tensor introduced in \cite{Bergshoeff:2014pca}. The parameter $s$ is sign, $\gamma$, $\bar{\sigma}$ and $\bar{\Lambda}_{0}$ are the parameters which defined in terms of other parameters like $\sigma, m$ and $\mu$.\\
The metric (\ref{eqmetric}) is a solution for the field equations (\ref{eqfield}) under the condition
\begin{equation}
\lambda=-\dfrac{4\sigma \mu^2 m^2+\gamma m^2-s\mu^2}{4\mu^2 m^2}.
\end{equation}
The Lagrangian (\ref{eqlag}) can be written as follows
\begin{equation}
L_{GMMG}=-\left(\sigma+\dfrac{c_{f}}{m^2}\right)e.R+\dfrac{1}{2\mu}L_{CS}+c_{h}e.T+\left[\dfrac{\Lambda_{0}}{6}-\dfrac{c_{f}^2}{2m^2}+\dfrac{\alpha c_{h}^2}{2}\right](e.e\times e),
\end{equation}
where we have used
\begin{equation}\label{eqqmf}
{h}=c_h {e},\;\;\; {f}=c_f {e}.
\end{equation}
The dreibein components of the metric after Wick rotation can be chosen as
\begin{equation}
e^{0}=\dfrac{d\xi}{\xi},\;\;\;\;\;\;\; e^{1}=\dfrac{\xi}{2}\left(dz-d\bar{z}\right),\;\;\;\;\; e^{2}=\dfrac{\xi}{2}\left(dz+d\bar{z}\right).
\end{equation}
Then, the spin connections would be
\begin{equation}
\omega^{1}_{0}=\dfrac{1}{2}\xi (dz-d\bar{z}),\;\;\;\; \omega^{2}_{0}=\dfrac{1}{2}\xi \left(dz+d\bar{z}\right),
\end{equation}
and therefore the dualized spin-connections are given by
\begin{equation}
\omega^{0}=0,\;\;\;\; \omega^{1}=\dfrac{\xi}{4}\left(dz+d\bar{z}\right),\;\;\;\; \omega^{2}=\dfrac{\xi}{4}\left(d\bar{z}-dz\right).
\end{equation}
The different terms of the action are given by
\begin{equation}\label{eqq32}
e.R=\dfrac{\xi}{4}d\xi \wedge dz \wedge d\bar{z},\;\;\;\;\;\; e.e\times e=\dfrac{\xi}{2}d\xi \wedge dz \wedge d\bar{z},\;\;\;\;\; T=0,\;\;\;\;\; L_{CS}=0,\;\;\;\;\; \omega .e=-\dfrac{\xi^2}{2}dz \wedge d\bar{z}.
\end{equation}
Then the on-shell action is given by
\begin{align}
&S_{GMMG}=\nonumber\\
&\dfrac{1}{4}\left(-\sigma +\dfrac{\Lambda_{0}}{3}-\dfrac{c_{f}^2}{m^2}+\alpha c_{h}^2-\dfrac{c_{f}}{m^2}\right)\left[\dfrac{1}{\delta^2}\iint e^{\phi}dzd\bar{z}+\dfrac{1}{2}\iint\psi dzd\bar{z}+\dfrac{\delta^2}{16}\iint e^{-\phi}\psi^2 dzd\bar{z}\right].
\end{align}
The boundary actions are given as
\begin{align}
S_{GH}&=\int -\left(\sigma +\dfrac{c_{f}}{m^2}\right) \omega .e+\dfrac{1}{2\mu} \omega .\omega +c_{h}e. e=\nonumber\\
&\dfrac{1}{2}\left(\sigma +\dfrac{c_{f}}{m^2}\right)\left[\dfrac{1}{\delta^2}\iint e^{\phi}dzd\bar{z}+\dfrac{1}{2}\iint\psi dzd\bar{z}+\dfrac{\delta^2}{16}\iint e^{-\phi}\psi^2 dzd\bar{z}\right] ,\\
S_{ct}&= -\dfrac{1}{4}\left(\sigma +\dfrac{c_{f}}{m^2}\right) \int \;e\;(1+\delta^2 \kappa(z,\bar{z})).
\end{align}
{If $\frac{\Lambda_{0}}{3}-\frac{c_{f}^2}{m^2}+\alpha c_{h}^2=0$, then we have}
\begin{align}
&S_{r}=-\dfrac{1}{8}\left(\sigma +\dfrac{c_{f}}{m^2}\right)\left[\iint\psi dzd\bar{z}+\dfrac{\delta^2}{8}\iint e^{-\phi}\psi^2 dzd\bar{z}\right]
\end{align}
%\begin{align}
%\lim \limits_{\substack{%
   % \mu \to \infty\\
    %\alpha \to 0}} c_{f}&= m\left(\sqrt{m^2\sigma^2+\Lambda_{0}}+\sigma m\right)\;\;\;\;\text{for} \;\;\;\;\sigma<0\\
    %\lim \limits_{\substack{%
    %m \to \infty\\
    %\alpha \to 0}} c_{f}&=2\mu^2\sigma^2-\dfrac{\Lambda_{0}}{2\sigma},\;\;\;\;\text{for} \;\;\;\;\sigma<0
 %\end{align}
%\begin{equation}
%\left(\sigma +\dfrac{c_{f}}{m^2}\right)=2\sigma\pm\sqrt{\sigma^2+\dfrac{\Lambda_{0}}{m^2}},\;\;\; \mu\to \infty ,\;\;\;\;\; \alpha\to 0
%\end{equation}
by using \eqref{eqq23}, one can obtain
\begin{equation}
S_{EE}=-\dfrac{1}{8}\left(\sigma +\dfrac{c_{f}}{m^2}\right)\left(\dfrac{1-n^2}{n}\right)\left[\log\left(\dfrac{\ell}{\delta}\right)+\dfrac{\delta^2}{n\ell^2}\right],
\end{equation}
one finally arrives at
\begin{equation}\label{eqqent50}
S_{EE}=\dfrac{c^{\prime}}{3}\log\left(\sqrt{\dfrac{24\pi}{\varsigma c^{\prime}}}\ell\right)+\dfrac{\varsigma c^{\prime 2}}{72 \pi \ell^2},\;\;\;\;c^{\prime}=\dfrac{c_{+}+c_{-}}{2}
\end{equation}
where
\begin{equation}
\varsigma=-\dfrac{64Gm^2\delta^2}{c_{f}+\sigma m^2},\;\;\;\;c_{\pm}=-\dfrac{3}{2G}\left(\sigma +\dfrac{\alpha c_{h}}{\mu}+\dfrac{c_{f}}{m^2}\pm \dfrac{1}{\mu}\right).
\end{equation}
The dreibein components of metric (\ref{eqqpon}) can be written as
\begin{equation}\label{eqq50}
e^{0}=\dfrac{r}{\eta}d\tau,\;\;\;\;\;\; e^{1}=\dfrac{n}{\eta}dr,\;\;\;\;\;\; e^{2}=\dfrac{1}{\eta}d\eta.
\end{equation}
The dualized spin connections would be
\begin{equation}\label{eqq51}
\omega^{0}=-\dfrac{n}{2\eta}dr,\;\;\;\;\;\; \omega^{1}=\dfrac{r}{2\eta}d\tau,\;\;\;\;\;\;\omega^{2}=\dfrac{1}{2n}d\tau.
\end{equation}
The presymplectic form of GMMG is given by \cite{Setare:2016jba}
\begin{equation}\label{eqthetagmmg}
\Theta_{GMMG}=-\left(\sigma +\dfrac{c_{f}}{m^2}\right)\delta \omega .e+\dfrac{1}{2\mu}\delta \omega .\omega +c_{h}\delta e. e.
\end{equation}
Then, we assume $\delta \omega =\partial \omega/\partial n$, $\delta e =\partial e/\partial n$, one can get
\begin{equation}
\Theta_{GMMG}=-\left(\sigma +\dfrac{c_{f}}{m^2}\right)\left[\dfrac{r}{2\eta^2}dr\wedge d\tau-\dfrac{1}{2n^2 \eta}d\tau\wedge d\eta\right].
\end{equation}
So, the entropy for GMMG is as follows
\begin{equation}\label{eqentropyG}
S=\int_{0}^{2\pi} d\tau \int \sqrt{-g}\Theta^{r}dz=\int_{0}^{2\pi} d\tau \int_{\eta_{f}}^{\ell} \dfrac{1}{2 n^2}\left(\sigma +\dfrac{c_{f}}{m^2}\right) \dfrac{d\eta}{\eta}=\dfrac{1}{4G}\left(\sigma +\dfrac{c_{f}}{m^2}\right)\ln\left(\dfrac{\ell}{\eta_{f}}\right).
\end{equation}
here $\eta_{f}=1/\xi_{f}$.
In the case of $\delta\ll 1$, one can obtain
\begin{equation}
S_{HEE}= \lim \limits_{\substack{%
    n \to 1\\
    r \to 0}}\dfrac{1}{4G}\left(\sigma +\dfrac{c_{f}}{m^2}\right) \left[\ln\left(\dfrac{\ell e^{\frac{\phi}{2}}}{\delta}\right)+\dfrac{\psi}{4e^{\phi}}\delta^2\right].
\end{equation}
After using \eqref{eqq23} and limiting, one can arrive \eqref{eqqent50}.

\subsection*{Entanglement Entropy for EGMG}
Exotic general massive gravity is the exotic version of NMG which is a parity$-$odd theory describing a propagating massive spin$-$2 fields in three dimensions. The Lagrangian of the theory in terms of auxiliary fields $f$ and $h$ is \cite{Ozkan:2018cxj}
\begin{align}
L_{EGMG}=&-\dfrac{1}{m^2}[f.R(\omega)+\dfrac{1}{6m^4}f.f\times f-\dfrac{1}{2 m^2}f.D(\omega)f+\dfrac{\nu}{2}f.e\times e-m^2h.T(\omega)+\nonumber\\
&\dfrac{(\nu-m^2)}{2}\left(\omega.d\omega+\dfrac{1}{3}\omega.\omega\times \omega\right)+\dfrac{\nu m^4}{3\mu}e.e\times e],\;\;\;\;\nu= 1-\dfrac{m^4}{\mu^2}.
\end{align}
The field equation of the theory in the metric formalism is obtained as follows
\begin{equation}\label{eqq1}
R_{\mu \nu}-\dfrac{1}{2}g_{\mu \nu}R+\Lambda g_{\mu \nu}+\dfrac{1}{\mu}C_{\mu \nu}-\dfrac{1}{m^{2}}H_{\mu \nu}+\dfrac{1}{m^{4}}L_{\mu \nu}= 0,
\end{equation}
where
\begin{equation}\label{eqq2}
H_{\mu \nu}=\epsilon_{\mu}^{\alpha \beta}\nabla_{\alpha}C_{\nu \beta},\hspace{0.5cm}L_{\mu \nu}=\dfrac{1}{2}\epsilon_{\mu}^{\alpha \beta}\epsilon_{\nu}^{\gamma \sigma}C_{\alpha \gamma}C_{\beta \sigma}.
\end{equation}
$\mu$ and $m$ are mass parameters and $H_{\mu \nu}$ and $L_{\mu \nu}$ are traceless and symmetric tensors.
The above Lagrangian can be rewritten by using (\ref{eqqmf}) as
\begin{equation}
L_{EGMG}=-\dfrac{ c_{f}}{m^2}e.R+\left(c_{h}+\dfrac{c_{f}^2}{2 m^4}\right)(e.De)-\dfrac{1}{m^2}(\nu -m^2)L_{CS}-\dfrac{1}{m^2}\left[\dfrac{c_{f}^3}{6m^4}+\dfrac{\nu c_{f}}{2}+\dfrac{\nu m^4}{3\mu}\right](e.e\times e).
\end{equation}
By using \eqref{eqq32}, one can obtain
\begin{equation}
S_{EGMG}=\dfrac{m^2}{8\mu}\iint \xi_{f}^2 dzd\bar{z}
\end{equation}
The GH term for EGMG is given by
\begin{equation}\label{eqboundray}
L_{GH}=-\dfrac{c_{f}}{2m^2}e.\omega +\dfrac{1}{2}\left(c_{h}+\dfrac{c_{f}^2}{m^4}\right)e.e+\left(1-\dfrac{\nu}{m^2}\right)\omega .\omega ,
\end{equation}
then the GH action is given
\begin{equation}
S_{GH}=-\dfrac{m^2}{4\mu}\iint \xi_{f}^2 dzd\bar{z},
\end{equation}
where we have used
\begin{equation}\label{eqchcf1}
c_{h}=\dfrac{1}{2}\left(1-\dfrac{1}{m^2}\right)\left(1-\dfrac{m^4}{\mu^2}\right),\;\;\;c_{f}=-\dfrac{m^4}{\mu}.
\end{equation}
The counter term is given
\begin{equation}
S_{ct}=\dfrac{m^2}{8\mu}\iint e(1+\delta^2\kappa(z,\bar{z})) dzd\bar{z}.
\end{equation}
The renormalized on-shell action by using cut-off surface is given as
\begin{align}
&S_{r}=\dfrac{m^2}{8\mu}\left[\iint\psi dzd\bar{z}+\dfrac{\delta^2}{8}\iint e^{-\phi}\psi^2 dzd\bar{z}\right],
\end{align}
then, similar to the previous section, using \eqref{eqq23} we have
\begin{equation}
S_{r}=-\dfrac{m^2}{8\mu}\left(\dfrac{1-n^2}{n}\right)\left[\log\left(\dfrac{\ell}{\delta}\right)+\dfrac{\delta^2}{n\ell^2}\right]
\end{equation}
one finally arrives at
\begin{equation}\label{eqq67}
S_{r}=\dfrac{c^{\prime}}{3}\log\left(\sqrt{\dfrac{24\pi}{\varsigma c^{\prime}}}\ell\right)+\dfrac{\varsigma c^{\prime 2}}{72 \pi \ell^2},\;\;\;\;c^{\prime}=\dfrac{c_{+}+c_{-}}{2}
\end{equation}
where
\begin{equation}
\varsigma=-\dfrac{64G\mu\delta^2}{m^2},\;\;\;\;\;\; c_{\pm}=\dfrac{3}{2G}\left[-\dfrac{m^2}{\mu}\pm \left(1+\dfrac{m^2}{\mu^2}-\dfrac{1}{m^2}\right)\right].
\end{equation}
The presymplectic form of EGMG is given \cite{Setare:2016jba}
\begin{equation}\label{eqprysemplecticform}
\Theta_{EGMG}=\dfrac{1}{2}\left[\left(c_{h}+\dfrac{c_{f}^2}{m^4}\right)\delta e.e+\left(1-\dfrac{\varpi}{m^2}\right)\delta \omega .\omega -\dfrac{c_{f}}{m^2}\delta e.\omega \right],
\end{equation}
this presymplectic using (\ref{eqq50}), (\ref{eqq51}) can be written as
\begin{equation}
\Theta=\dfrac{c_{f}}{2m^2}\left[\dfrac{r}{2\eta^2}dr\wedge d\tau+\dfrac{1}{2n^2 \eta}d\tau\wedge d\eta\right].
\end{equation}
Then, one can obtain the entropy for EGMG as follows
\begin{equation}\label{eqEGMGHEE}
S=\int_{0}^{2\pi} d\tau \int \sqrt{-g}\Theta^{r}d\eta=\int_{0}^{2\pi} d\tau \int_{\eta_{f}}^{\ell} \dfrac{ c_{f}}{4 n^2 m^2}\dfrac{d\eta}{\eta}=
-\dfrac{m^2}{16G \mu}\ln\left(\dfrac{\ell}{\eta_{f}}\right),
\end{equation}
here $\eta_{f}=1/\xi_{f}$. In the case of $\delta\ll 1$, one can obtain
\begin{equation}
S_{HEE}= -\lim \limits_{\substack{%
    n \to 1\\
    r \to 0}}\dfrac{m^2}{16G \mu} \left[\ln\left(\dfrac{\ell e^{\frac{\phi}{2}}}{\delta}\right)+\dfrac{\psi}{4e^{\phi}}\delta^2\right].
\end{equation}
After using \eqref{eqq23} and limiting, one can arrive \eqref{eqq67}.

\section{Conclusion}\label{sec4}

In this paper we have calculated the entanglement entropy of $T\bar{T}$-deformed CFT to  leading order correction to the entropy due to the finite cut-off of AdS$_{3}$ in Poincare coordinates in the framework of NMG, GMMG and EGMG theories.
We obtained the on-shell action of each theory and expanded it up to leading order of cut-off of spacetimes. This correction term is correspond to the deformation which comes from the bulk term. We have also obtained the entropy for the deformed theories using Song's method \cite{Song:2016pwx}. We find an agreement between the results in the two methods.\\

{\bf Acknowledgements}\\
We thank Amin Faraji for reading the article and for his encouraging.\\
The authors acknowledge the support of Kurdistan University.

\appendix

\end{document}